\documentclass[twocolumn,showpacs,aps,pra,amsmath,amssymb,superscriptaddress]{revtex4-1}

\usepackage{graphicx}
\usepackage{dcolumn}
\usepackage{bm}
\usepackage{hyperref}
\usepackage{subfigure}
\usepackage{bbding}

\usepackage{multirow}
\usepackage{makecell}
\usepackage{diagbox}
\usepackage{float}

\usepackage{epstopdf}
\usepackage{tikz}
\usepackage{color,bbm}

\newcommand{\beq}{\begin{equation}}
\newcommand{\eeq}{\end{equation}}
\newcommand{\bqa}{\begin{eqnarray}}
\newcommand{\eqa}{\end{eqnarray}}
\newcommand{\nn}{\nonumber}

\newcommand{\rt}[1]{\sqrt{#1}\,}

\newcommand{\bra}[1]{ \langle{#1} |}
\newcommand{\ket}[1]{ |{#1} \rangle}

\newcommand{\sq}[1]{\left[ {#1} \right]}

\newcommand{\an}[1]{\left\langle{#1}\right\rangle}
\newcommand{\tr}[1]{{\rm Tr}\sq{ {#1} }}

\newtheorem{Theorem}{Theorem}
\newtheorem{Definition}{Definition}
\newtheorem{Corollary}{Corollary}
\newtheorem{Lemma}{Lemma}

\definecolor{maroon}{rgb}{0.7,0,0}

\definecolor{ngreen}{rgb}{0.3,0.7,0.3}

\definecolor{golden}{rgb}{0.8,0.6,0.1}

\begin{document}

\title{Polygon relations and subadditivity of entropic measures for discrete and continuous multipartite entanglement}

\author{Lijun Liu}
\affiliation{College of Mathematics and Computer Science, Shanxi Normal University, Linfen 041000, China}

\author{Xiaozhen Ge}
\email{gexiaozhen18@gmail.com}
\affiliation{Shanghai Institute of Intelligent Science and Technology, Tongji University, Shanghai 201804, China}
\affiliation{Department of Applied Mathematics, The Hong Kong Polytechnic University, Kowloon 999077, Hong Kong, China}

\author{Shuming Cheng}
\email{drshuming.cheng@gmail.com}
\affiliation{The Department of Control Science and Engineering, Tongji University, Shanghai 201804, China}
\affiliation{Shanghai Institute of Intelligent Science and Technology, Tongji University, Shanghai 201804, China}
\affiliation{Institute for Advanced Study, Tongji University, Shanghai, 200092, China}

\begin{abstract}
	
	In a recent work by us [Ge {\it et al., Phys. Rev. A 110, L010402 (2024)}], we have derived a series of polygon relations of bipartite entanglement measures that is useful to reveal entanglement properties of discrete, continuous, and even hybrid multipartite quantum systems. In this work, with the information-theoretical measures of R\'enyi and Tsallis entropies, we study the relationship between the polygon relation and the subadditivity of entropy. In particular, the entropy-polygon relations are derived for pure multi-qubit states and then generalized to multi-mode Gaussian states, by utilizing the known results from the quantum marginal problem. Then the equivalence between the polygon relation and subadditivity is established, in the sense that for all discrete or continuous multipartite states, the polygon relation holds if and only if the underlying entropy is subadditive. As a byproduct, the subadditivity of R\'enyi and Tsallis entropies is proven for all bipartite Gaussian states. Finally, the difference between polygon relations and monogamy relations is clarified, and generalizations of our results are discussed. Our work provides a better understanding of the rich structure of multipartite states, and hence is expected to be helpful for the study of multipartite entanglement.
	
\end{abstract}

\thanks{Corresponding authors: X. Ge and S. Cheng}

\maketitle

\section{Introduction}

Multipartite entanglement is of significant importance in utilizing quantum information technology, including quantum internet~\cite{Kimble2008} and quantum computer~\cite{Ladd2010}, to revolutionize our daily life. It has generated a great interest in its characterization and quantification, which, however, is a challenging task, due to the complicated structures of multipartite states and operations~\cite{Horodecki2009,Weedbrook2012,Eric2014,Szalay2015}. In a recent work by us~\cite{Ge2023B}, a set of polygon relations, based on some well-explored entanglement measures, is obtained as a geometric tool to reveal the entanglement properties of multipartite quantum states. Importantly, it provides a unified approach to faithfully quantifying genuine tripartite entanglement of discrete, continuous, and even hybrid systems, rather than certifying the presence of entanglement via a well-chosen witness~\cite{Guhne2009} or Fisher information matrix~\cite{Gessner2016}.

In this work, we delve further into the study of polygon relations of multipartite entanglement in discrete and continuous systems based on R\'enyi~\cite{Renyi1961} and Tsallis~\cite{Tsallis1988} entropies that also measure the coherence~\cite{Streltsov2017,Cheng2015,Rastegin2016,Che2023} and uncertainty~\cite{Luo2005B,Luo2005A,Hall2023} of the state. With these information-theoretical measures, we examine the connection between the polygon relation and the subadditivity of entropy, and also clarify the difference between polygon relations and monogamy relations for multipartite entanglement.

Particularly, the polygon relation, in the form of $\mathcal{E}_{A_i|\bar{A}_i}\leq \sum_{j\neq i}\mathcal{E}_{A_j|\bar{A}_j}$, dictates that for any pure $N$-partite state shared by parties $A_1, \dots, A_N$, the degree of entanglement between one single party $A_i$ and its rest $\bar{A_i}\equiv A_1\dots A_N\backslash A_i$, quantified by the measure $\mathcal{E}$, is less than the sum of entanglement shared by any other party $A_j\neq A_i$ and the rest $\bar{A}_j$~\cite{Qian2018,Yang2022,Ge2023B}. As the bipartition entanglement $\mathcal{E}_{A_i|\bar{A}_i}$ is typically determined by the local reduced state $\rho_{A_i}$, the polygon relation is closely related to the quantum marginal problem of whether there exists a physical state with reduced states $\rho_{A_i}$~\cite{Higuchi2003,Klyachko2006,Eisert2008,Schilling2014,Tyc2015}. Indeed, following the known results from the marginal problem~\cite{Higuchi2003,Eisert2008}, we are able to derive the polygon relations obtained in~\cite{Qian2018,Ge2023B} for pure multi-qubit states and obtain those of R\'enyi and Tsallis entropies for pure multipartite Gaussian states. 

Noting further that any bipartite state of the discrete system with a finite dimension can be purified to a pure tripartite state, we establish the equivalence between the polygon relation and the subadditivity of entropy that the former holds if and only if the underlying entropy is subadditive, i.e. $E(\rho_{A_iA_j})\leq E(\rho_{A_i})+E(\rho_{A_j})$ for any bipartite state $\rho_{A_iA_j}$ and entropy $E$. This connection is then generalized to the infinite-dimensional system, and thus the subadditivity of R\'enyi and Tsallis entropies is obtained for Gaussian states as a byproduct. It has its own interest that the subadditivity of entropy implies the nonnegativity of quantum mutual information~\cite{Adami1997,Nielsen2000,Gorisman2005} and the relevant research mainly focuses on the finite-dimensional system~\cite{Araki1970,Auden2007,Linden2013,Zhu2015,Chehade2019}. It is also interesting to note that the Tsallis entropy is subadditive in both discrete and Gaussian systems, while the R\'enyi entropy is not for qudits~\cite{Linden2013}.

Finally,  it is pointed out that the entropy-polygon relations can be directly utilized as a geometric tool introduced in~\cite{Xie2021,Ge2023,Ge2023B} to reveal genuine multipartite entanglement, and is complementary to the monogamy relation for entanglement~\cite{Coffman2000,Terhal2004,Dhar2017}. 

This work is structured as follows. Sec.~\ref{sec2} presents a basic introduction to entropies of both discrete and continuous states, including von Neumann, R\'enyi, and Tsallis entropies, and recaps the subadditivity of entropy, the entropy-polygon relation, and the quantum marginal problem. In Sec.~\ref{sec3}, the entropy-polygon relation is derived for all pure multipartite states that each party is a single qubit or a single Gaussian mode.  Sec.~\ref{sec4} establishes the equivalence between the polygon relation and the subadditivity of entropy for general multipartite states and proves that the R\'enyi and Tsallis entropies are subadditive for all bipartite Gaussian states. Discussions and conclusions are given in Sec.~\ref{sec5}. 

\section{Entropic measures for discrete and continuous quantum states} \label{sec2}

In this section, a brief introduction to the entropy of quantum states in discrete and continuous systems, including the celebrated von Neumann entropy and its two generalizations--R\'enyi and Tsallis entropies, is first presented in Sec.~\ref{quantumentropy}. Then, the subadditivity of entropy for bipartite states and the polygon relation of entropic measures for multipartite states are recapped in Secs.~\ref{subadditivity of entropy} and~\ref{polygonrelationentropy}, respectively. Finally, some known results about the quantum marginal problem are presented in Sec.~\ref{quantum marginal}.

\subsection{Quantum entropy}\label{quantumentropy}

The entropy of a quantum state, measuring the information or uncertainty contained in the quantum system, is crucial in the theory of quantum information. In analogy to Shannon entropy in the classical regime, the von Neumann entropy of state $\rho$ is defined by
\beq
	S(\rho)=-\tr{\rho\log\rho}.\label{von}
\eeq
The logarithm base is left arbitrary here and elsewhere, corresponding to a choice of units. It has many mathematical implications and physical insights in quantum information theory, ranging from the Holevo bound~\cite{Holevo1973} to quantum communication protocols~\cite{Devetak2004,Horodecki2005}. Another important classes of quantum entropy are the R\'{e}nyi entropy~\cite{Renyi1961}
\beq
	R_p(\rho)=\frac{\log \tr{\rho^p}}{1-p}, ~~p>1,\label{renyi}
\eeq
and the Tsallis entropy~\cite{Tsallis1988}
\beq
	T_q(\rho)=\frac{1-\tr{\rho^q}}{q-1},~~q>1. \label{tsallis}
\eeq
In the limit $p\rightarrow 1$ and $q\rightarrow 1$, both the R\'enyi and Tsallis entropies recover the von Neumann entropy~(\ref{von}) and thus can be considered as a generalization of the latter. 

All above entropies are nonnegative and remain invariant under the isometry, i.e., $E(\rho)=E(U\rho U^\dag)\geq 0$ for any isometry $U$ and entropy $E$. This immediately yields that quantum entropy of an  density matrix  $\rho$ is completely determined by its eigenvalues
\beq
{\rm spec}(\rho)=\{\lambda_i\}, \label{eigenvalue}
\eeq
with the spectrum decomposition $\rho=\sum_i\lambda_i\ket{\lambda_i}\bra{\lambda_i}$. For example, the state of a discrete quantum system with the Hilbert space of finite dimension $d$ admits
	\beq
	S(\rho)=-\sum_{i=1}^d \lambda_i\log \lambda_i,
\eeq
and
	\beq
			\tr{\rho^x}=\sum_{i=1}^d \lambda_i^x\label{trhod}
	\eeq
for the R\'{e}nyi entropy with $x=p$ and the Tsallis entropy with $x=q$. 

By contrast, the continuous system has an infinite state dimension, and it is convenient to introduce the quadrature phase operators $\hat{p}$ and $\hat{q}$ to characterize its state via the covariance matrix (CM). Especially, any $n$-mode Gaussian state can be described by a $ 2n\times 2n$ CM $\sigma$, with elements $\sigma_{ij}=\an{{{\hat\zeta }_i}{{\hat \zeta }_j} + {{\hat \zeta }_j}{{\hat \zeta }_i}} - \left\langle {{{\hat \zeta }_i}} \right\rangle \left\langle {{{\hat \zeta }_j}} \right\rangle$ where $\an{\bullet}={\rm tr}({\rho\,\bullet})$ and $\hat{\zeta} = (\hat q_1, \hat p_1, \dots, \hat q_n, \hat p_n)^{\top}$ is the vector of the phase operators. Besides, an isometry $U$ acting on $\rho$ is translated into a real symplectic operation $V$ on $\sigma$, i.e., $\rho^\prime=U\rho U^\dag\rightarrow \sigma^\prime=V\sigma V^\top$, and the spectrum decomposition of $\rho$ corresponds to the symplectic spectrum decomposition $V\sigma V^\dag={\rm diag} (s_1, s_1,\dots, s_n, s_n)$, with symplectic eigenvalues~\cite{Adesso2014}
\beq
{\rm sspec}(\rho)\equiv{\rm sspec}(\sigma)=\{s_i\}. \label{symplecticvalue}
\eeq
Correspondingly, there is the von Neumann entropy~\cite{Adesso2004}
\beq
S(\rho)=\sum_i\left[\frac{s_i+1}{2}\log \frac{s_i+1}{2}-\frac{s_i-1}{2}\log\frac{s_i-1}{2}\right], \label{vonGaussian}
\eeq
and 
\beq
\tr{\rho^x}=\prod_{i=1}^n g_x(s_i) \label{rhoGaussian}
\eeq
for the R\'{e}nyi and Tsallis entropies, with~\cite{Holevo2000}
\beq
g_x(y)=\frac{2^x}{(y+1)^x-(y-1)^x}. \label{gfunction}
\eeq
If $x=2$, then it follows from Eqs.~(\ref{rhoGaussian}) and~(\ref{gfunction}) that $\tr{\rho^2}=\prod_{i=1}^n 1/s_i=1/\rt{\det(\sigma)}$, implying the state purity of $\rho$ is determined by the determinant of $\sigma$.

\subsection{Subadditivity of entropy} \label{subadditivity of entropy}

Suppose that a pure state $\ket{\psi}_{AB}$, either discrete or continuous, is shared by two parties denoted by $A$ and $B$. The entropies of its two reduced states $\rho_{A(B)}={\rm Tr}_{B(A)}\left[\ket{\psi}_{AB}\bra{\psi}\right]$ are identical to each other
\beq
	E(\rho_A)=E(\rho_B) \label{identity}
\end{equation}
for $E$ being von Neumann, R\'enyi, or Tsallis entropy. Since any pure state has zero entropy, it is easy to find $E(\ket{\psi}_{AB}) \leq E(\rho_A)+E(\rho_B)$. 

Generally, if 
\begin{equation}
	E(\rho_{AB})\leq E(\rho_A)+E(\rho_B)\label{Subadditivity}
\end{equation}
holds for an arbitrary bipartite state $\rho_{AB}$ with reduced states $\rho_{A(B)}={\rm Tr}_{B(A)}\left[\rho_{AB}\right]$, then $E$ is said to be subadditive. Indeed, the subadditivity has been proven for von Neumann entropy~\cite{Araki1970}, Tsallis entropy in the discrete system~\cite{Auden2007} and Gaussian system with order $q=2$~\cite{Ge2023B}, and R\'enyi entropy in the Gaussian system with order $p=2$~\cite{Adesso2012,Lami2016}. However, the subadditivity may not always be satisfied, and one such example to violate the subadditivity is the R\'enyi entropy in general discrete systems~\cite{Linden2013}. 

It is remarked that the subadditivity~(\ref{Subadditivity}) indicates that the quantum mutual information between parties $A$ and $B$, defined by $I(A, B):=E(\rho_A)+E(\rho_B)-E(\rho_{AB})$, is always non-negative, and hence is expected to be a reasonable quantity which could measure the correlations between $A$ and $B$~\cite{Adami1997,Devetak2004}.

\subsection{Polygon relation of entanglement measures}\label{polygonrelationentropy}

For a bipartite pure state $\ket{\psi}_{AB}$, the entropy of its local reduced states as per Eq.~(\ref{identity}), is also an entanglement measure $\mathcal{E}$~\cite{Vidal2000}. Any such measure $\mathcal{E}$ defined on pure states can be extended to mixed states $\rho_{AB}$ via the convex-roof construction~\cite{Horodecki2009}
\beq
\mathcal{E}(\rho_{AB}):={\rm inf}_{\{p_m,\ket{\psi^{(m)}}_{AB}\}}\sum_m p_m\,\mathcal{E}(\ket{\psi^{(m)}}_{AB}),
\eeq
where the infimum is over all pure decompositions $\rho_{AB}=\sum_m p_m\ket{\psi^{(m)}}_{AB}\bra{\psi^{(m)}}_{AB}$. It is noted that except for the pure product states, the entanglement measure of a bipartite state is generally not equivalent to its entropy, i.e.,
\beq
 \mathcal{E}(\rho_{AB}) \neq E(\rho_{AB}). \label{distinction}
 \eeq
This distinction can be further drawn from the difference between polygon relations and monogamy relations of entropic measures, which will be discussed in Sec.~\ref{sec5}. 

When it comes to the $N$-partite state $\ket{\psi}_{A_1\cdots A_N}$ shared by parties $A_1, \dots, A_N$ with $N\geq3$, there are $N$ one-to-other bipartitions: $A_1|\bar{A}_1, \cdots, A_N|\bar{A}_{N}$. If $\mathcal{E}$ is an entanglement measure, then it is easy to obtain
\beq
\mathcal{E}_{A_i|\bar{A}_i}\equiv\mathcal{E}(\ket{\psi}_{A_i|\bar{A}_i})=E(\rho_{A_i})=E(\rho_{\bar{A}_i}),~~\forall i=1,\cdots,N.
	\eeq
 Furthermore, it has been shown that these $\mathcal{E}_{A_i|\bar{A}_i}$ satisfy the polygon relation~\cite{Qian2018,Ge2023B}
\beq
	\mathcal{E}_{A_i|\bar{A}_i}\leq \sum_{j\neq i} \mathcal{E}_{A_j|\bar{A}_j},~~\forall i=1,\cdots,N,\label{Polygon}
\eeq
for some well-chosen measures $\mathcal{E}$, such as the von Neumann and Tsallis entropies. This relation suggests that the bipartite entanglement between one single party and its rest is always less than the sum of entanglement shared by any other party and the corresponding rest.

\subsection{Quantum marginal problem }\label{quantum marginal}

Both the subadditivity of entropy~(\ref{Subadditivity}) and the polygon relation of entanglement measures~(\ref{Polygon}) are essentially determined by the reduced states. This is closely related to the quantum marginal problem of whether the given local states are compatible with a global physical state. There have been a few useful results obtained in the marginal problem, which will be shown to be useful in studying the polygon relation of the subadditivity.

For example, there must be~\cite{Higuchi2003}
\begin{equation}
	\lambda_{A_i}\leq \sum_{j\neq i}\lambda_{A_j}\label{nqubit}
\end{equation}
for any pure $N$-qubit state $\ket{\psi}_{A_1\cdots A_N}$,  where $\lambda_i$ denotes the smallest eigenvalue of $\rho_{A_i}$. Similarly, any pure $n$-mode Gaussian state obeys~\cite{Eisert2008}
\beq
	s_{A_i}-1\leq \sum_{j\neq i} (s_{A_j}-1),\label{nGaussianmode}
\eeq
and a general $n$-mode Gaussian state satisfies~\cite{Eisert2008}
\begin{eqnarray}
	\sum_{j=1}^n d_j&\geq& \sum_{j=1}^n s_j, ~~\forall k=1,\cdots, n \label{gaussianmarginal}\\
	d_n-\sum_{j=1}^{n-1}d_j&\leq& s_n-\sum_{j=1}^{n-1} s_j.\label{nmode}
\end{eqnarray}
Here, the symplectic eigenvalues $s_i$ and diagonal elements $d_i$ of the CM $\sigma$ are arranged in non-decreasing order, with ${\rm diag}(\sigma)={\rm diag}(d_1, d_1, d_2, d_2,\dots, d_n, d_n )$.

\section{Polygon relations for pure multi-qubit and -mode Gaussian states } \label{sec3}

 Here we utilize the above results from the quantum marginal problem to derive the entropy-polygon relations for the pure multi-partite state $\ket{\psi}_{A_1\cdots A_N}$ that each party $A_i$ has one single qubit or one single Gaussian mode. Before proceeding to prove this claim, we first introduce the definition of a monotone function.

\begin{Definition}\label{definition1}
	A function $f(x): \mathbb{R}\rightarrow \mathbb{R}$, is monotonic if it is non-decreasing and concave under the parameter $x$.
\end{Definition}

It is obvious that any smooth function $f(x)$, with the first-order derivative $f^\prime(x)\geq 0$ and the second-order $f^{\prime\prime}(x)\leq 0$, is monotonic. The monotone function admits the following property:

\begin{Lemma}\label{lemma1}
	Assume that $f(x)$ is a monotone function with $f(0)=0$. If $x_i\leq \sum_{j\neq i}x_j$ for a nonnegative number set $\{ x_i\}$, then $f(x_i)\leq \sum_{j\neq i}f(x_j)$.
\end{Lemma}

The proof of Lemma~\ref{lemma1} is deferred to Appendix~\ref{prooflemma}. In the following, we use this Lemma to derive the polygon relation as desired.

\subsection{One single qubit per party} \label{marginal}

If the local reduced state is a qubit, it has two eigenvalues $\{\lambda, 1-\lambda\}$ with $\lambda\equiv \min \{\lambda_1, \lambda_2\}\in [0, 1/2]$. As a consequence, its entropies simplify to
\begin{align}
	f_S(\lambda)=& -\lambda\log \lambda-(1-\lambda)\log (1-\lambda), \label{von1}\\
	f_{R_p}(\lambda)=& \frac{\log[\lambda^p+(1-\lambda)^p]}{1-p},\label{ren1}\\
	f_{T_q}(\lambda)=& \frac{1-[\lambda^q+(1-\lambda)^q]}{q-1}.\label{tsallis1}
\end{align}
It is further proven in Appendix~\ref{monotonic} that the von Neumann entropy $S$, the R\'{e}nyi entropy $R_p$ with  $p\leq 2$, and the Tsallis entropy $T_q$ are monotone functions of $\lambda$.
 
 Recall that for any pure $N$-qubit state $\ket{\psi}_{A_1\cdots A_N}$, the smallest eigenvalues of local reduced states satisfy~(\ref{nqubit}).  Following Lemma~\ref{lemma1} immediately yields 
 \beq
 f_{E=S, R_p, T_q}(\lambda_{A_i})\leq \sum_{j\neq i} f_{E=S, R_p, T_q}(\lambda_{A_j}) \label{polygonfunction}
 \eeq
 for the entropic functions~(\ref{von1})-(\ref{tsallis1}). Then, substituting the equalities $\mathcal{E}_{A_i|\bar{A}_i}= E(\rho_{A_i})=f_E(\lambda_{A_i})$ for all $i$ into Eq.~(\ref{polygonfunction}) gives rise to the desired polygon relation 
\begin{equation}
	\mathcal{E}_{A_i|\bar{A}_i}\leq \sum_{j\neq i}\mathcal{E}_{A_j|\bar{A}_j}. \label{poly1}
\end{equation}
The above result encompasses previous ones obtained in~\cite{Zhu2015,Qian2018,Ge2023B}.

It is also found that the polygon relation in terms of the R\'{e}nyi entropy with $p>2$ does not hold generally in the $n$-qubit system. Consider a family of W-class states
\beq
\ket{W}=a_1\ket{10\cdots 0}+a_2\ket{010\cdots 0}+\cdots+a_N\ket{0\cdots01} \nn
\eeq
with $a_1^2\geq 1/2$ and $a_i\neq0$ for any $i$. As $\lambda_{A_1}=1-a_1^2$ and $\lambda_{A_k}=a_k^2$ for all $k=2,\cdots,N$, it is easy to obtain the equality $\lambda_{A_1}=\sum_{k=2}^N \lambda_{A_k}$. We can further show that $f_{R_p}(\lambda)$ is strictly convex in some interval $[0,\delta_\alpha]$ for $p>2$, due to $f_{R_p}''(0)=p/(p-1)>0$. Thus, the polygon relation is violated as
\begin{align}
	&f_{R_p}(\lambda_{A_1})=f_{R_p}(\sum_{j\neq1} \lambda_{A_j}) \nn \\
	=&\sum_{j\neq1}\left[ \frac{\lambda_{A_j}}{\sum_{j\neq1}\lambda_{A_j}}f_{R_p}(\sum_{j\neq1} \lambda_{A_j})+\left(1-\frac{\sum_{j\neq 1 }\lambda_{A_j}}{\sum_{j\neq1}\lambda_{A_j}}\right)f_{R_p}(0)\right]\nn \\
    >& \sum_{j\neq1}f_{R_p}(\lambda_{A_j}). \nn
\end{align}
Here we use $ f(0)=0$ and the strict convexity that $f(xy_1+(1-x)y_2)< xf(y_1)+(1-x)f(y_2)$ for any $x\in(0,1)$.

\subsection{Single Gaussian mode per party}

 The entropies of a single-mode Gaussian state are determined by its unique symplectic eigenvalue $s$ as
\begin{align}
		g_S(s)&= \frac{s+1}{2}\log\left(\frac{s+1}{2}\right)-\frac{s-1}{2}\log\left(\frac{s-1}{2}\right),\label{von2} \\
	     g_{R_p}(s)&= \frac{\log[(s+1)^p-(s-1)^p]-p}{p-1},\label{ren2}\\
	g_{T_q}(s)&=\frac{1}{q-1}\left[1-\frac{2^q}{(s+1)^q-(s-1)^q}\right].\label{Tsallis2}
\end{align}
It is shown in Appendix~\ref{monotonic} that all above functions are monotonic. Let $\tilde s=s-1$, and it is evident that the function $g_{E}(\tilde s):=g_{E}(s-1)$ is also a monotone function of $\tilde s\geq0$, with $g_{E}(0)=0$ for $E=S, R_p, T_q$.

Since the symplectic eigenvalues of local reduced states satisfy~(\ref{nGaussianmode}) for any pure $N$-mode Gaussian state, it again follows directly from Lemma~\ref{lemma1} that 
\begin{eqnarray}
	g_{E=S,R_p,T_q}(\tilde s_{A_i})\leq \sum_{j\neq i}g_{E=S,R_p,T_q}(\tilde s_{A_j}).
\end{eqnarray}
 Replacing $g_{E}(\tilde s_{A_i})=E(\rho_{A_i})$ with $\mathcal{E}_{A_i|\bar{A}_i}$ immediately leads to the polygon relations~(\ref{Polygon}) of von Neumann, R\'enyi, and Tsallis entropies for any pure $N$-mode Gaussian state.

\section{Equivalence between the polygon relation and subadditivity of entropy}\label{sec4}

 For pure $3$-qubit and $3$-mode states, combining the polygon relation~(\ref{Polygon}) with the equation~(\ref{identity}) leads to
\begin{align}
	E(\rho_{A_iA_j})&=E(\rho_{A_k})=\mathcal{E}_{A_k|\bar{A}_k} \nn \\
	&\leq \mathcal{E}_{A_i|\bar{A}_i}+\mathcal{E}_{A_j|\bar{A}_j}  \nn \\
	&=E(\rho_{A_i})+E(\rho_{A_j}) \label{inverse}
\end{align}
for $E$ being von Neumann, R\'enyi, or Tsallis entropy. It equals to the subadditivity of the underlying entropy in the special case where the bipartite state is reduced from a pure tripartite one.

 In turn, if the subadditivity of quantum entropy~(\ref{Subadditivity}) is valid, then we are able to derive the corresponding polygon relation by noting
\begin{align*}
	\mathcal{E}_{A_N|\bar{A}_N}&=E(\rho_{A_1\cdots A_{N-1}})\leq E(\rho_{A_1\cdots A_{n-2}})+ E(\rho_{A_{N-1}}) \nn \\
	&\leq \cdots \leq \sum_{i=1}^{N-1} E(\rho_{A_i})=\sum_{i=1}^{N-1}\mathcal{E}_{A_i|\bar{A}_i}.	\label{iteration}
	\end{align*}
It indicates that the subadditivity of entropy is sufficient for the entropy-polygon relation. Then, we show that the polygon relation is also sufficient for the subadditivity, hence establishing the equivalence between them.

\begin{Theorem}\label{Theorem1}
	For all pure multipartite discrete and/or Gaussian states, the entropy-polygon relation as per~(\ref{Polygon}) is valid if and only if the corresponding entropy is subadditive. 
\end{Theorem} 

The {\it if}~part follows directly from Eq.~(\ref{iteration}), and the {\it only if}~part from combining Eq.~(\ref{inverse}) with the state-purification technique that it is always possible to purify any bipartite (discrete and Gaussian) state to a pure tripartite state~\cite{Holevo2001}. 

We remark that Theorem~\ref{Theorem1} holds in a general context that all pure multipartite states, either discrete or continuous, are required. If this requirement is relaxed, then Theorem~\ref{Theorem1} may not hold and hence the equivalence between the polygon relation and the subadditivity could be broken. For example, it has been shown in Sec.~\ref{marginal} that the polygon relation, based on the R\'enyi entropy with $p=2$, is valid for all pure multi-qubit states, however, the corresponding subadditivity can be violated by qubits~\cite{Adesso2003}.

Next, we show that the entropy-polygon relations do hold in both the discrete and Gaussian systems, and both R\'enyi and Tsallis entropies are subadditive for all Gaussian states, thus significantly generalizing previous results obtained in~\cite{Adesso2012,Lami2016,Ge2023B}.

\begin{table*}[!htbp]
	\renewcommand{\arraystretch}{1.2}
	\centering
	\setlength{\tabcolsep}{4mm}
	\begin{tabular}{|p{2.8cm}<{\centering}|p{0.5cm}<{\centering}|p{2.08cm}<{\centering}|p{2.08cm}<{\centering}|p{2.08cm}<{\centering}|p{2.08cm}<{\centering}|}
		\hline 
		\multicolumn{2}{|c|}{\multirow{2}{*}{\diagbox[width=5cm]{Property}{System}} }& \multicolumn{2}{ c|}{Discrete} & \multicolumn{2}{c|}{Continuous}\\
		\cline{3-6}
		\multicolumn{2}{|c|}{~}& qubit & qudit & Gaussian & non-Gaussian\\
		\hline	
		\multirow{6}{*}{Subadditivity~(\ref{Subadditivity})} & \multirow{2}{*}{$S$~(\ref{von})} & \multirow{2}{*}{\Checkmark~\cite{Araki1970}} &\multirow{2}{*}{\Checkmark~\cite{Araki1970}} & \multirow{2}{*}{\Checkmark~\cite{Lieb1973}}& \multirow{2}{*}{\Checkmark~\cite{Lieb1973}} \\
		& & & & &\\
		\cline{2-6}
		& \multirow{2}{*}{$R_p$~(\ref{renyi})} & \multirow{2}{*}{\XSolidBrush~\cite{Adesso2003}} & 
		\multirow{2}{*}{\XSolidBrush~\cite{Linden2013}} & \Checkmark~$p=2$~\cite{Adesso2012} &\multirow{2}{*}{\XSolidBrush~\cite{Linden2013}}\\
		& & & & \Checkmark~$p>1$~Ours & \\
		\cline{2-6}
		&\multirow{2}{*}{$T_q$~(\ref{tsallis})} & \multirow{2}{*}{\Checkmark~\cite{Auden2007}} & \multirow{2}{*}{\Checkmark~\cite{Auden2007}} & \multirow{2}{*}{\Checkmark Ours} & \multirow{2}{*}{\textbf{?}}\\
		& & & & &\\
		\hline	
		\multirow{6}{*}{Polygon Relation~(\ref{Polygon})} & \multirow{2}{*}{$S$~(\ref{von})} & \multirow{2}{*}{\Checkmark~\cite{Qian2018}} &\multirow{2}{*}{\Checkmark~\cite{Ge2023B}} & \multirow{2}{*}{\Checkmark Ours}& \multirow{2}{*}{\Checkmark~\cite{Lieb1973}} \\
		& & & & &\\
		\cline{2-6}
		& \multirow{2}{*}{$R_p$~(\ref{renyi})} &\Checkmark~$p\leq2$~Ours& 
		\multirow{2}{*}{\XSolidBrush~Ours} & \multirow{2}{*}{\Checkmark~Ours}&\multirow{2}{*}{\XSolidBrush~Ours}\\
		& &\XSolidBrush~$p>2$~Ours & &  & \\
		\cline{2-6}
		&\multirow{2}{*}{$T_q$~(\ref{tsallis})} & \multirow{2}{*}{\Checkmark~\cite{Ge2023B}} & \multirow{2}{*}{\Checkmark~\cite{Ge2023B}} & \Checkmark~ $q=2$~\cite{Ge2023B} & \multirow{2}{*}{\textbf{?}}\\
		& & & & \Checkmark~$q>1$~Ours&\\
		\hline
	\end{tabular}
	\caption{A summary of the entropic properties in different quantum systems. With the von Neumann~(\ref{von}), R\'enyi~(\ref{renyi}), and Tsallis~(\ref{tsallis}) entropies, the problems of whether they are subadditive~(\ref{Subadditivity}) and satisfy the polygon relation~(\ref{Polygon}) are investigated. \Checkmark represents the validity of the property, \XSolidBrush~indicates that there exist examples to invalidate this property, and the question mark $\textbf{?}$ means it still leaves open. Theorem~\ref{Theorem1} establishes the equivalence between the subadditivity and the polygon relation for general qudits and multi-mode Gaussian states, which, however, can be broken when the system is restricted to qubits with R\'enyi entropy. It is interesting to note that the Tsallis entropy is subadditive in both discrete and Gaussian systems, while the R\'enyi entropy is only in Gaussian systems as proven in~Theorem~\ref{Theorem2}. Since the qudit state is regarded as a class of non-Gaussian states, some results about the non-Gaussian case are directly borrowed from those in the qudit case. } 
	\label{Tab1}
\end{table*}

\subsection{A qudit beyond qubit per party}\label{qudit}

In the discrete quantum system with a finite dimensional Hilbert space, both the von Neumann entropy and Tsallis entropy are subadditive~\cite{Araki1970,Auden2007}, whereas the R\'enyi entropy is not~\cite{Linden2013}.  Correspondingly, the polygon relations of the von Neumann and Tsallis entropies can be naturally obtained via Theorem~\ref{Theorem1} for all pure multipartite states.

\subsection{Multiple Gaussian modes per party}\label{multiGaussianmode}

Any $m$-mode Gaussian state $\sigma$ has $m$ symplectic eigenvalues $s_i$, denoted by $\boldsymbol{s}=(s_1, s_2, \dots, s_m)^\top$. It follows first from Eqs.~(\ref{vonGaussian})-(\ref{gfunction}) that its entropies are
\begin{eqnarray}
	S(\rho)&\equiv &G_S(\boldsymbol{s})=G_{S}(s_1,\cdots,s_m)= \sum_{i=1}^m g_S(s_i), \label{multivon}\\
	R_p(\rho)&\equiv& G_{R_p}(\boldsymbol{s})=G_{R_p}(s_1,\cdots,s_m)=\sum_{i=1}^m g_{R_p}(s_i),\label{multiren}\\
	T_q(\rho)&\equiv& G_{T_q}(\boldsymbol{s})=G_{T_q}(s_1,\cdots,s_m)=\frac{[1-\prod_{i=1}^mg_q(s_i)]}{q-1}, \nonumber 
\end{eqnarray}
where the functions $g_S$ and $g_{R_p}$ are from Eqs.~(\ref{von2}) and~(\ref{ren2}), representing the von Neumann and R\'enyi entropy associated with $s_i$ respectively. 

We then introduce the notation of vector majorization.

\begin{Definition}\label{definition2}
	For two vectors $\boldsymbol{x},\boldsymbol{y}\in \mathbb{R}^d$ with $x_{(i)}$ $(y_{(i)})$ being the $i$-th smallest element of $\boldsymbol{x}$ $(\boldsymbol{y})$, if $\sum_{i=1}^kx_{(i)}\geq\sum_{i}^ky_{(i)}$ for $k=1,\cdots,d-1$ and $\sum_{i=1}^d x_{(i)}=\sum_{i}^d y_{(i)}$, then $\boldsymbol{x}$ is said to be majorized by $\boldsymbol{y}$, written as $\boldsymbol{x}\prec\boldsymbol{y}$. If $\sum_{i=1}^kx_{(i)}\geq\sum_{i}^ky_{(i)}$ for all $k=1,\cdots,d$, then $\boldsymbol{x}$ is weakly majorized by $\boldsymbol{y}$, written as $\boldsymbol{x}\prec_w\boldsymbol{y}$.
\end{Definition}

Then, there is:

\begin{Lemma}~\cite{Marshall1979} \label{Lemma2}
For every concave function $g$, there is $\sum_i g(x_{(i)})\geq \sum_ig(y_{(i)})$ for $\boldsymbol{x}\prec\boldsymbol{y}$. Furthermore, if $g$ is also non-decreasing, then  $\sum_i g(x_{(i)})\geq \sum_ig(y_{(i)})$ for $\boldsymbol{x}\prec_w\boldsymbol{y}$.
\end{Lemma}

It is evident from Eq.~(\ref{gaussianmarginal}) that  the vector $\boldsymbol{d}$, describing the main diagonal elements of the $m$-mode Gaussian CM $\sigma$, is weakly majorized by its symplectic-eigenvalue vector $\boldsymbol{s}$. Thus, it is readily to obtain the following result:

\begin{Theorem}\label{Theorem2}
	The polygon relation as per~$(\ref{Polygon})$, in terms of von Neumann, R\'enyi, and Tsallis entropies, is valid for any pure multipartite Gaussian state.
\end{Theorem}

The proof of Theorem~\ref{Theorem2} is as follows. For an arbitrary pure $N$-partite Gaussian state $\ket{\psi}_{A_1\cdots A_N}$, there is
\begin{equation}
	\mathcal{E}_{A_N|\bar{A}_{N-1}}=E(\rho_{A_1\cdots A_{N-1}})=G_{E=S,R_p,T_q}(\boldsymbol{s}),
\end{equation}
where $\boldsymbol{s}={\rm sspec}(\rho_{A_1\cdots A_{N-1}})=\{ s_i \}$. It is always possible to choose a proper symplectic operation $V_{A_1}\oplus \cdots \oplus V_{A_{N-1}}$ to diagonalize the CM $\sigma$ of the reduced Gaussian state $\rho_{A_1\cdots A_{N-1}}$ such that
\begin{align} \label{sigmaprime}
  \sigma^\prime=& V_{A_1}\oplus  \cdots  \oplus V_{A_{N-1}} \sigma V^\top{A_1}\oplus \cdots \oplus V^\top_{A_{N-1}} \nn \\
	=&\left(
	\begin{array}{cccc}
		\sigma_{A_1} & \gamma_{A_1A_2} &\cdots & \gamma_{A_1A_{N-1}}\\
		\gamma_{A_1A_2} &\sigma_{A_2} & \cdots & \gamma_{A_2A_{N-1}}\\
		\vdots & \vdots& \ddots &\vdots\\
		\gamma_{A_{1}A_{N-1}} & \gamma_{A_2A_{N-1}}&\cdots & \sigma_{A_{N-1}}
	\end{array}
	\right),
\end{align}
where $\gamma_{A_iA_j}$ represents the correlation matrix between parties $A_i$ and $A_j$ and the CM of each reduced $m_i$-mode Gaussian state $\rho_{A_i}$ is in a diagonal form $\sigma_{A_i}={\rm diag}( d^i_1, d^i_1\cdots,d^i_{m_i},d^i_{m_i})$, with symplectic eigenvalues ${\rm sspec}(\rho_{A_i})=\{d^i_k  \}_{k=1}^{m_i}$.  Denote the diagonal vector of $\sigma^\prime$ by $\boldsymbol{d}^\prime:=(d^i_j: j=1,\cdots,m_i, i=1,\cdots,n-1 )$. Then, it follows from Eq.~(\ref{gaussianmarginal}) and Definition~\ref{definition2} that
\beq
\boldsymbol{d}^\prime \prec_w, \boldsymbol{s}
\eeq
and consequently, using Lemma~\ref{Lemma2} gives rise to
\beq
\sum_l^{\sum_j m_j} g(d_{(l)}) \geq \sum_l^{\sum_j m_j} g(s_{(l)}), \label{chain}
\eeq
for any monotone function $g$. Note that both the function $g_S$ for the von Neumann entropy~(\ref{von2}) and $g_{R_p}$ for the R\'enyi entropy~(\ref{ren2}) are nondecreasing and concave (See Appendix~\ref{monotonic}), and thus Eq.~(\ref{chain}) can be written as
\begin{align}
\mathcal{E}_{A_n|\bar{A}_n}&=G_{E=S,R_p}(\boldsymbol{s})\leq G_{E=S,R_p}(\boldsymbol{d}^\prime) \nn \\
&=\sum_{i=1}^{N-1}\sum_{j=1}^{m_i}g_{S,R_p}(d_j^i)=\sum_{i=1}^{N-1} G_{E=S,R_p}(\boldsymbol{d}_i) \nn \\
&=\sum_{i=1}^{N-1} E(\rho_{A_i})=\sum_{i=1}^{N-1} \mathcal{E}_{A_i|\bar{A}_i}.\label{ineq1}
\end{align}
The equalities about $G_{E=S, R_p}$ are directly from Eqs.~(\ref{multivon}) and~(\ref{multiren}). With respect to the Tsallis entropy, we have
\begin{align}
	&\sum_{i=1}^{N-1} \mathcal{E}_{A_i|\bar{A}_i}-\mathcal{E}_{A_n|\bar{A}_n} \nn \\
	=&\sum_{i=1}^{N-1} \frac{1-\prod_{j=1}^{m_i}g_q(d_j^i)}{q-1}-\frac{1- [\prod_{l=1}^{\sum_j m_j}g_q(s_l)]}{q-1} \nn \\
	\geq& \frac{1-\prod_{i=1}^{N-1} [\prod_{j=1}^{m_i}g_q(d_j^i)]}{q-1}-\frac{1- [\prod_{l=1}^{\sum_j m_j}g_q(s_l)]}{q-1} \nn \\
	=&\frac{\prod_{l=1}^{\sum_j m_j}g_q(s_l)-\prod_{i=1}^{N-1} [\prod_{j=1}^{m_i}g_q(d_j^i)]}{q-1} \nn \\
	\equiv& \frac{H(\boldsymbol{s})-H(\boldsymbol{d}^\prime)}{q-1}
	\end{align}
The first equality follows from Eq.~(\ref{multitsa}), and the inequality from $1-\prod_i x_i\leq \sum_i (1-x_i)$ for any $x_i\in[0,1]$. Since the function $\log H$ equals to the R\'enyi entropy with $p=q$ multiplying a negative number $1-q$, we can obtain $H(\boldsymbol{s})\geq H(\boldsymbol{d}^\prime)$ from $G_{R_q}(\boldsymbol{s})\leq G_{R_q}(\boldsymbol{d}^\prime)$ as shown in Eq.~(\ref{ineq1}), hence proving the polygon relation in terms of Tsallis entropy and completing the proof of Theorem~\ref{Theorem2}.

The above results about the entropy-polygon relation immediately lead to the property of subadditivity.

\begin{Corollary}\label{corollary}
	The von Neumann, R\'enyi, and Tsallis entropies are subadditive for bipartite Gaussian states. 
\end{Corollary} 

This significantly generalizes the subadditivity of R\'enyi entropy from order $p=2$~\cite{Adesso2012,Lami2016} to $p>1$, and of Tsallis entropy from order $q=2$~\cite{Ge2023B} to $q>1$. Finally, all relevant results about the entropy polygon relations and the subadditivity of entropy are summarized in Table~\ref{Tab1}.

\section{Discussion}\label{sec5}

The entropy-polygon relation~(\ref{Polygon}) not only suggests that entanglement shared between one single party and its rest is always less than the sum of entanglement shared by other one-to-other bipartitions, but also provides a geometric picture that the bipartite entanglement, quantified by an entropic measure, can be interpreted as the side of a polygon~\cite{Qian2018}. This geometric interpretation can be perfectly illustrated via tripartite entanglement that for any pure tripartite state, its bipartite entanglements correspond to sides of a triangle. Furthermore, it is shown in~\cite{Ge2023B} that the triangle has nonzero area if and only if the underlying state is genuinely entangled,  and the modified area is a faithful measure for genuine tripartite entanglement in the sense that it is monotonic under (Gaussian) local operations and classical communication. Hence, it provides a unified geometric tool to reveal multipartite entanglement in the discrete and continuous quantum systems. 

Monogamy of entanglement, dictating that it cannot be freely shared between arbitrarily many parties, admits a form of~\cite{Coffman2000}
\beq
\mathcal{E}(\rho_{A_iA_j})+\mathcal{E}(\rho_{A_iA_k}) \leq \mathcal{E}(\rho_{A_i|\bar{A}_i})
\eeq
for a tripartite quantum state $\ket{\psi}_{A_iA_jA_k}$ and some proper entanglement measure $\mathcal{E}$. If $\mathcal{E}$ is chosen as the entropic measures as discussed in this work, then it follows from the entropy-polygon relation~(\ref{Polygon}) that 
\begin{align}\label{relation}
	\mathcal{E}(\rho_{A_iA_j})+\mathcal{E}(\rho_{A_iA_k}) &\leq \mathcal{E}(\rho_{A_i|\bar{A}_i}) \nn \\
	& \leq\mathcal{E}(\rho_{A_j|\bar{A}_j})+\mathcal{E}(\rho_{A_k|\bar{A}_k}) \nn \\
	&= E(\rho_{A_j})+E(\rho_{A_k})  \nn \\
	&= E(\rho_{A_iA_j})+E(\rho_{A_iA_k}).
	\end{align}
The first and second inequalities imply that the polygon relation is complementary to the monogamy relation that  the former upper bounds $\mathcal{E}(\rho_{A_i|\bar{A}_i})$ and the latter lower bounds it. Further, the first inequality and the second equality signals the distinction between $\mathcal{E}(\rho)$ and $E(\rho)$ given in Eq.~(\ref{distinction}). For example, we consider the three-qubit  GHZ state $\ket{{\rm GHZ}}=(\ket{000}+\ket{111})/\sqrt{2}$, with reduced states $\rho_{AB}=\rho_{AC}=\rho_{BC}=\frac{1}{2}(|00\rangle\langle 00|+|11\rangle\langle 11|)$. It is easy to verify that $\mathcal{E}(\rho_{AB})=\mathcal{E}(\rho_{AC})=0$ for any entanglement measure $\mathcal{E}$, whereas $E(\rho_{AB})=E(\rho_{AC})>0$ for any entropic measure $E$, hence confirming the validity of Eq.~(\ref{relation}) and signaling the difference between entanglement measure and entropic measure.

We also point out that it is possible to apply Lemma~\ref{lemma1} to further generalize the entropy-polygon relation~(\ref{Polygon}) to 
\beq
f(\mathcal{E}_{A_i|\bar{A}_i})\leq \sum_{j\neq i} f(\mathcal{E}_{A_j|\bar{A}_j})~\label{generalisation}
\eeq
for any quantity $f(\mathcal{E})$ that is defined as a monotone function of $\mathcal{E}$. For example, it is known that the Tsallis entropy is a monotone function of the R\'enyi entropy, i.e., $T_q(\rho)=f(R_q(\rho))$ with $f(x)=[e^{(1-q)x}-1]/(1-q)$. Thus, it follows immediately from the above equation that the Tsallis-polygon relation can also be derived from the R\'enyi-polygon one. It is interesting to investigate whether other commonly-used entropies, such as unified entropy~\cite{Hu2006}, admit the polygon relation or not, and partial results have been obtained in~\cite{Yang2022}.

We have borrowed the known results from the quantum marginal problem to derive the entropy-polygon relation and the subadditivity of entropy, especially for the single-qubit and single-mode Gaussian cases in Sec.~\ref{sec3}. It is remarked that the polygon relation or the subadditivity can be considered as the constraints imposed on the reduced states, which may shed new light on the quantum marginal problem in the general qudit or multi-mode Gaussian cases. It is also interesting to study the problem of whether those properties or constraints are applied to the most general non-Gaussian system.

Finally, our results are expected to aid significant progress in the study of discrete, continuous, and even hybrid multipartite entanglement,  and the methods used in this work are also hoped to find applications in studying other quantum resources, such as genuine nonlocality~\cite{Tavakoli2022} and steering~\cite{Xiang2022}.

\begin{acknowledgments} 
	This research is supported by  the Shanghai Municipal Science and Technology Fundamental Project (No. 21JC1405400), the National Natural Science Foundation of China (Nos. 12205219 and 62173288), Innovation Program for Quantum Science and Technology 2023ZD0300600, Guangdong Provincial Quantum Science Strategic Initiative (No. GDZX2200001), and Hong Kong Research Grant Council (No. 15213924).
\end{acknowledgments}

\appendix 

\section{Proof of Lemma~\ref{lemma1}}\label{prooflemma}

The proof Lemma~\ref{lemma1} in the main text is given as follows. Following first that the monotone function is non-decreasing, we have 
\begin{equation}
	f(x_i)\leq  f(\sum_{j\neq i}^m x_j) \label{nondecresing}
\end{equation}
for $0\leq x_i \leq \sum_{j\neq i} x_j$. Further, we can obtain
\begin{eqnarray}
	&&f(\sum_{j\neq i} x_j )\nn\\
	&=&\sum_{j\neq i} \left[\frac{x_j}{\sum_{j\neq i}x_j }f(\sum_{j\neq i} x_j)+\left(1-\frac{x_j}{\sum_{j\neq i}x_j }\right) f(0)\right]\nn \\
	&\leq & \sum_{j\neq i} f(x_j), \label{monotonicity}
\end{eqnarray}
where $ f(0)=0$ is used and the inequality follows from the concavity that $f(py_1+(1-p)y_2)\geq pf(y_1)+(1-p)f(y_2)$ for any $p\in[0,1]$. Combining Eqs.~(\ref{nondecresing}) with~(\ref{monotonicity}) completes the proof.

\section{ Monotonicity of the entropic functions}\label{monotonic}

It follows from Eqs.~(\ref{von1})-(\ref{tsallis1}) in the main text that after direct computation, we have

\begin{align}
	f_{S}'(\lambda)=& \log(1-\lambda)-\log\lambda \geq 0,\\
	f_{S}''(\lambda)=&-\left[\frac{1}{\lambda}+\frac{1}{1-\lambda}\right]\leq 0, \\
f_{R_p}'(\lambda)=& \frac{p[(1-\lambda)^{p-1}-\lambda^{p-1}]}{(p-1)[\lambda^p+(1-\lambda)^p]}\geq 0,\\
	f_{R_p}''(\lambda)=&\frac{p[(1-\lambda)^{p-1}-\lambda^{p-1}]^2}{(p-1)[\lambda^p+(1-\lambda)^p]^2}\nn\\
	&-\frac{p(p-1)\lambda^{p-2}(1-\lambda)^{p-2}}{(p-1)[\lambda^p+(1-\lambda)^p]^2} \nn \\
	\leq&  0, \\
		f_{T_q}'(\lambda)=&\frac{q[(1-\lambda)^{q-1}-\lambda^{q-1}]}{q-1}\geq 0,\\
f_{T_q}''(\lambda)=&- q[(1-\lambda)^{q-2}+\lambda^{q-2}]\leq 0,
\end{align}
for $\lambda\in[0,1/2]$, $p \in (1, 2]$, and $q>1$. Thus, these entropic functions are non-decreasing  and concave under $\lambda$. It is also interesting to note that for the R\'{e}nyi function with $p>2$, there is $f_{R_p}''(0)=p/(p-1)>0$. Due to the continuity of the function $f_{R_p}''$, there must exist $\lambda_\alpha$ such that $f_{R_p}''(\lambda)\geq0$ for any $\lambda\in[0,\lambda_\alpha]$, indicating that $f_{R_p}$ with $p>2$ is convex under $\lambda$ in this interval.

For the single-mode Gaussian state, the entropies can be written as 
\begin{align}
	g_{S}(s)&= \frac{s+2}{2}\log\left(\frac{s+2}{2}\right)-\frac{s}{2}\log\left(\frac{s}{2}\right), \\
    g_{R_p}(s)&= \frac{\log[(s+2)^p-s^p]-p}{p-1},\\
    g_{T_q}(s)&=\frac{1}{q-1}\left[1-\frac{2^q}{(s+2)^q-s^q}\right].
	\end{align}
Then we have
\begin{align}
	g_{S}'(s)&=\frac{1}{2}\log\left(1+\frac{2}{s}\right) \geq 0,\\
    g_{S}''(s)&=\frac{-1}{s(s+2)} \leq 0,\\ 
    	g_{R_p}'(s)&=\frac{p\,[(s+2)^{p-1}-s^{p-1}]}{(p-1)[(s+2)^p-s^p]} \geq 0,\\
   g_{R_p}''(s)&= \frac{-4p(p-1)(s+2)^{p-2}s^{p-2}}{(p-1)[(s+2)^p]-s^p]^2}\nn\\
   &~~~-\frac{p((s+2)^{p-1}-s^{p-1})^2}{(p-1)[(s+2)^p]-s^p]^2}\nn\\ 
    &\leq 0, \\
g_{T_q}'(s)&=\frac{2^q\, q\, [(s+2)^{q-1}-s^{q-1}]}{(q-1)[(s+2)^q-s^q]^2} \geq 0,\\
	g_{T_q}''(s)
	&=-\frac{2^{q+2} q (s+2)^{q-2}s^{q-2}}{[(s+2)^q]-s^q]^3}\nn\\
    &~~~-\frac{2^q q(q+1)[(s+2)^{p-2}-s^{p-1})^2]}{(q-1)[(s+2)^q]-s^q]^3} \nn \\
	&\leq 0,
\end{align}
for any non-negative $s$ and $p, q>1$. It follows above that they are monotonic under the parameter $s$.

\bibliography{Monotonicity2}

\end{document}